\documentclass[12pt,preprint]{aastex}

\shortauthors{J.J.M. in 't Zand et al.}
\shorttitle{Identification of counterparts to X-ray transient in NGC 6440}

\begin{document}

\slugcomment{Accepted for ApJL}
\title{Identification of the optical and quiescent counterparts to
the bright X-ray transient in NGC~6440}
\author{J.J.M.~in~'t~Zand\altaffilmark{1,2,3}, 
 M.H.~van~Kerkwijk\altaffilmark{1},
 D.~Pooley\altaffilmark{4},
 F.~Verbunt\altaffilmark{1},
 R.~Wijnands\altaffilmark{4,5},
 W.H.G.~Lewin\altaffilmark{4}}
\affil{}
\altaffiltext{1}{Astronomical Institute, Utrecht University, P.O. Box 80 000, 3508 TA Utrecht, the Netherlands}
\altaffiltext{2}{SRON National Institute for Space Research, 
	Sorbonnelaan 2, 3584 CA Utrecht, the Netherlands}
\altaffiltext{3}{email jeanz@sron.nl}
\altaffiltext{4}{Center for Space Research, Massachusetts Institute of Technology, Cambridge, MA 02139-4307}
\altaffiltext{5}{{\em Chandra\/} fellow}

\begin{abstract}
After 3 years of quiescence, the globular cluster NGC~6440 exhibited a bright
transient X-ray source turning on in August 2001, as noted with the RXTE
All-Sky Monitor. We carried out a short target of opportunity observation
with the {\em Chandra\/} X-ray Observatory and are able to associate the
transient with the brightest of 24 X-ray sources detected during quiescence
in July 2000 with {\em Chandra}. Furthermore, we securely identify the optical
counterpart and determine that the 1998 X-ray outburst in NGC~6440 was from
the same object. This is the first time that an optical counterpart to a
transient in a globular cluster is securely identified. Since the transient is
a type I X-ray burster, it is established that the compact accretor is a
neutron star. Thus, this transient provides an ideal case to study the
quiescent emission in the optical and X-ray of a transiently accreting neutron
star while knowing the distance and reddening accurately. One model that fits
the quiescent spectrum is an absorbed power law plus neutron star hydrogen
atmosphere model. We find an intrinsic neutron star radius of
$17_{-12}^{+31}$~km and an unabsorbed bolometric luminosity for the neutron
star atmosphere of $(2.1\pm0.8)\times10^{33}$~erg~s$^{-1}$ which is consistent
with predictions for a cooling neutron star.
\end{abstract}


%==============================================================================

\section{Introduction}
\label{secintro}
Currently, thirteen bright X-ray sources are known to inhabit twelve Galactic
globular clusters (for a recent overview, see Sidoli et al. 2001; the 13th source
was recently discovered by White \& Angelini 2001). They are low-mass
X-ray binaries (LMXBs) of which at least 11
harbor a neutron star. Four are transient in nature: those in Liller 1,
Terzan 5 and 6, and NGC 6440 (one could argue that Terzan~1 contains a
transient source as well with an activity lasting decades, see Guainazzi,
Parmar \& Oosterbroek 1999). Thus far, none of these transients had a
securely identified optical counterpart. The prospects for finding one
were best for the transient in NGC~6440 because it has a relatively low
reddening of $E(B-V)=1.07$ and a non-extreme distance of $8.5\pm0.4$~kpc
(Ortolani, Barbuy \& Bica 1994). The transient in NGC~6440 was seen in
outburst in Dec. 1971 (Markert et al. 1975; Forman, Jones \& Tananbaum
1976) and Aug. 1998 (In~'t~Zand et al. 1999; Verbunt et al. 2000). From ROSAT
HRI localizations of two sources in quiescence and optical observations during
the 1998 outburst, two candidate optical counterparts were identified,
V1 and V2 (Verbunt et al. 2000). V2 was evaluated to be the most likely one
with $B\approx22.7$ during outburst.

Pooley et al. (2001) observed NGC~6440 for 23 ksec with the {\em Chandra\/}
X-ray Observatory on July 4, 2000, when the transient was quiescent. They
found 24 point sources with 0.5-2.5 keV luminosities between 1.8$\times10^{30}$
and $9.9\times10^{32}$~erg~s$^{-1}$, which they designate CX1 through CX24.
The brightest source, CX1, coincides with the optical counterpart candidate V2.

The transient in NGC~6440 erupted again in August 2001, as detected with the
All-Sky Monitor (ASM) on the Rossi X-ray Timing Explorer RXTE\footnote{
as announced by the ASM team at {\tt http://xte.mit.edu}}.
We proceeded to secure the identification of the optical counterpart through
swift, high-resolution X-ray observations, and report the
results in the present paper. A preliminary report was already given by
In 't Zand et al. (2001).

%==============================================================================
\section{Observations and source identification}
\label{section}

The ASM (Levine et al. 1996; Bradt et al. 2000)
monitors NGC~6440 for 10 months of each year
at a detection threshold of about 15~mCrab in the 1.5 to 12 keV band per day of
observation. The transient exceeded this threshold during two weeks starting
on August 2, 2001. The peak flux was $60\pm10$~mCrab. Continued ASM monitoring
showed a reflaring episode starting on Sep. 10, with fluxes up to 0.2~Crab
and lasting for already almost 2 months at the time of this writing.
The Wide Field Cameras (WFCs) on BeppoSAX (Jager et al. 1997; In 't Zand 2001)
also observed NGC~6440 during this episode and detected at least 10 type-I
X-ray bursts. As soon as the community was alerted of the
outburst (on August 13), we proposed X-ray observations with {\em Chandra}
and optical observations with the European Southern Observatory's
New Technology Telescope in Director's Discretionary Time. 

\begin{figure}
\plotone{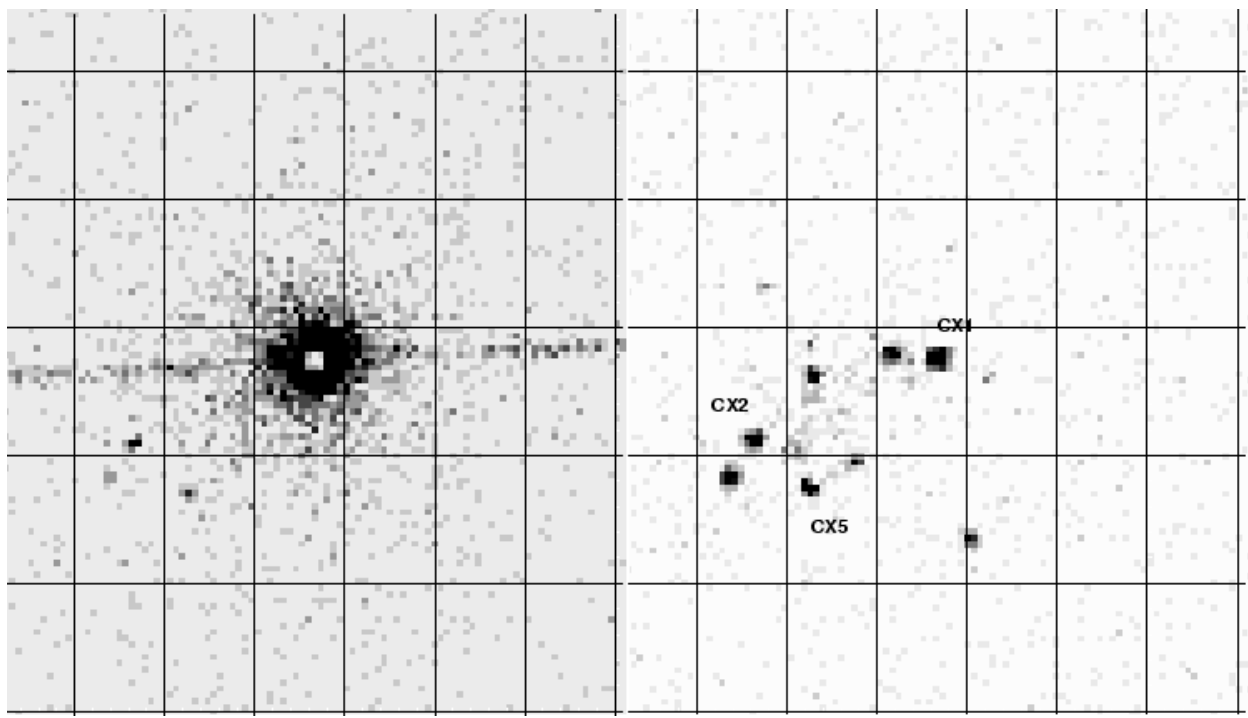}
  \caption{Chandra ACIS-S images of NGC~6440 during the transient active state
(left; 2.6 ks exposure) and quiescence (right; 23 ks exposure). The grids are
fixed to the equatorial coordinate system and defined in the same manner for
both cases. Pixels are 0\farcs5 on the side. The left panel shows the bright
transient with its large point spread function. The hole in the middle of the
PSF and the streak are artifacts introduced by the brightness of the transient.
The bright transient is coincident with quiescent source 'CX1' (Pooley et al.
2001). CX1 is the
brightest X-ray source in the globular cluster during quiescence with a
0.5-7 keV unabsorbed luminosity of $(1.5\pm0.5)\times10^{33}$~erg~s$^{-1}$.
\label{figchandra}
}
\end{figure}

The {\em Chandra\/} (Weiskopf, O'Dell \& van Speybroek 1996) 
target-of-opportunity observation (TOO) was approved and carried out on August
18.9 UT for an exposure time of 2.6~ksec with the Advanced CCD Imaging
Spectrometer (ACIS; Garmire, Nousek \& Bautz, in prep.) and no grating.
Despite the anticipated heavy pile up, we chose this configuration, and not
HRC, to optimize the sensitivity such that not only the transient but also a
few of the other sources in NGC~6440 would be detected (ACIS has 2 to 4 
times more effective area for 0.3-8 keV photons than HRC-I). Thus, the
association of the transient with one
of the quiescent sources would become independent from the {\em Chandra\/}
attitude accuracy. Only a $2\arcmin\times8\arcmin$ subarray of the
back-illuminated S3 chip was read so that the CCD frame time was 0.8~s instead
of 3.2~s and pile up effects were reduced. Figure~\ref{figchandra} shows
the obtained image, zoomed in on the globular cluster and compared to the
observation during quiescence (Pooley et al. 2001). It is evident that the
transient was active. Furthermore, it is closest to CX1, the brightest of the
sources detected during quiescence. Based on the {\em Chandra\/}
aspect solution for the 2001 observation (accurate to about 0.5\arcsec) and
the astrometry carried out by Pooley et al. (2001) for the 2000 observation,
the angular distance between the transient and CX1 is 0.14\arcsec\ (the
position of the transient is determined through weighing pixel position with
pixel values). Of the remaining sources, CX2 and CX5 are clearly detected
during the TOO as well, with photon rates in the full bandpass of
$(8\pm2)\times10^{-3}$ and $(6\pm1)\times10^{-3}$~s$^{-1}$. These rates
compare well with those during the earlier observation
($(7.6\pm0.6)\times10^{-3}$ and $(4.3\pm0.4)\times10^{-3}$~s$^{-1}$
respectively). The angular distances of the transient to CX2 and CX5 are
consistent with that to CX1 in the July 2000 image within 0.28$\pm$0.13\arcsec.
The distance is inconsistent with all of the other 23 sources from the July
2000 observation by at least 15$\sigma$. This confirms the transient-CX1
association.

A total of $2.6\times10^3$ events was detected within 8.0\arcsec\ from the
source centroid. In order to estimate the flux and avoid piled-up pixels,
we extracted a spectrum from the wings of the point spread
function (PSF) for pixels that are between 2.5 and 8.0\arcsec\ from the
centroid. We ignore negligible contributions from other sources in this
annulus. An absorbed power law fits the 0.7-8.0 keV spectral data well, with
a photon index of $1.2\pm0.1$, $N_{\rm H}=(8.2\pm1.6)\times10^{21}$~cm$^{-2}$
and $\chi^2_\nu=1.40$ ($\nu=75$ degrees of freedom).
After a correction for the fraction of the PSF covered (9\%), the flux is
$1.1\times10^{-10}$~erg~s$^{-1}$cm$^{-2}$ (0.5-10 keV) and the luminosity
$9\times10^{35}$~erg~s$^{-1}$ ($1.1\times10^{36}$~erg~s$^{-1}$ unabsorbed).
The statistical uncertainty is 7\%, that due to PSF uncertainties an estimated
10\%. The flux is roughly 10
times fainter than at the peak and $10^3$ times brighter than during
quiescence (see below). The source shows no strong
variability, with an upper limit of 20\%
at 90\% confidence for 100~s time scales. No narrow features are apparent in
the power density spectrum.

The position of CX1 (Pooley et al. 2001) is consistent with that of the
candidate optical counterpart V2 in Verbunt et al. (2000). They are
0.46\arcsec\ apart while the uncertainty in the optical position alone is
0.5\arcsec. We conclude that V2 is the secure optical counterpart to the
transient in NGC~6440, that CX1 is the quiescent X-ray counterpart, and that
the 1998 and 2001 outbursts were from the same object and not from any of the
other 3 or 4 transient low-mass X-ray binaries in quiescence that are likely
to be present in the cluster (Pooley et al. 2001). Most importantly, since
type-I X-ray bursts were detected from the transient during both 1998 and
2001 outbursts, it is now firmly established that CX1 is a quiescent neutron
star.

NGC~6440 was observed for us in Service Time with the New Technology Telescope
(NTT) at La Silla 3.5~d after the {\em Chandra\/} observation using the Superb
Seeing Imager SUSI2. The seeing was around 1\arcsec. Images were obtained in
the Bessell R and B filters. The analysis of these data is complicated
by proper motions of a number of stars in the field. However, the raw images
do show V2 to be active again in B at a level comparable to the 1998 outburst. 

%==============================================================================
\section{The quiescent X-ray spectrum}
\label{sectionspec}

\begin{figure}
\plotone{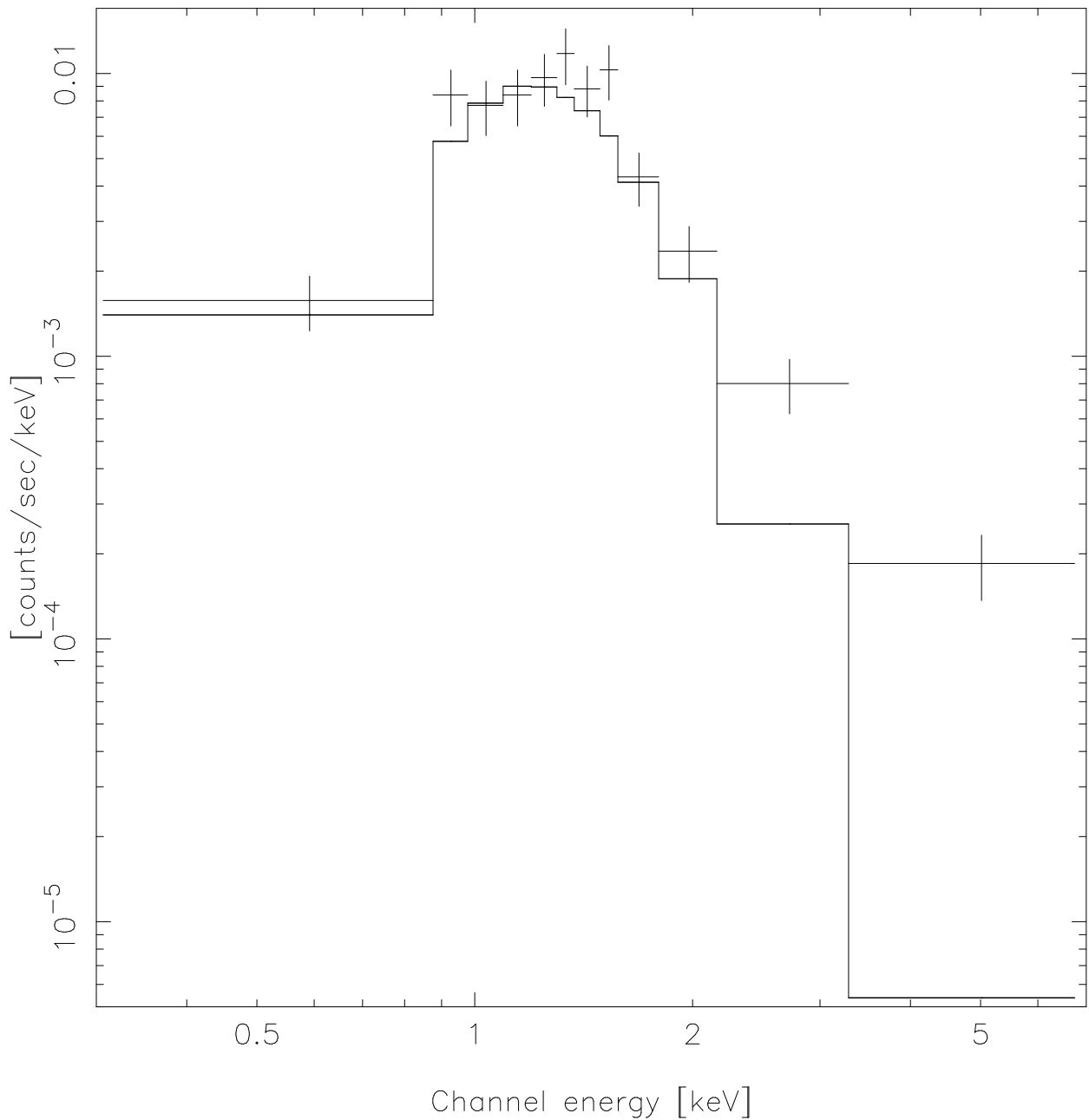}
  \caption{The spectrum of the quiescent emission of the
transient, as obtained in July 2000. The crosses refer to the measurements,
with the vertical lines indicating the errors, and the histogram refer to the 
fluxes per channel as predicted by the best-fit NS atmosphere model plus
power law, after resetting the power law contribution to zero to illustrate
its impact at high energies. The resetting also illustrates that both
spectral components cover fairly distinct photon energies.
\label{figspectrum}
}
\end{figure}

\begin{figure}
\plotone{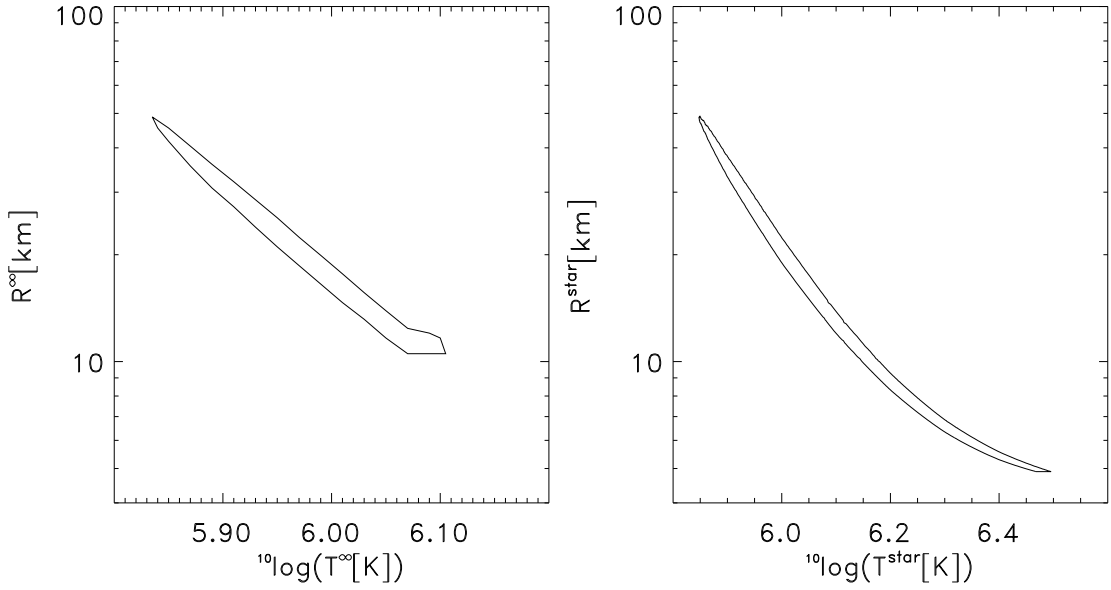}
  \caption{Contour plots of 90\%-confidence (R,T) region of NS atmosphere model
fit to CX1 spectrum during quiescence between 0.3 and 6.7 keV, for an observer
at infinity (left) or at the NS surface (right). It is assumed that the
distance is 8.5~kpc and
$N_{\rm H}=8.2\times10^{21}$~cm$^{-2}$. The location of the contour is
independent of NS mass for an observer at infinity.
\label{figcontour}
}
\end{figure}

Quiescent neutron stars (qNSs) exhibit X-ray emission at typical luminosities
of $10^{32}-10^{33}$~erg~s$^{-1}$ (e.g., Asai et al. 1998, McClintock 2001).
A plausible explanation for this emission is thermal radiation from the
atmosphere of a neutron star that is cooling from the previous outburst
(Brown, Bildsten \&
Rutledge 1998). We extracted the spectrum of the transient during quiescence
by accumulating all 251 photons within 1.5\arcsec\ from the source centroid and
applying background corrections as found elsewhere in the image (amounting
merely to a 2.3 photons subtraction). To ensure applicability of the $\chi^2$
statistic, we 
rebinned the data so that each channel contains at least 20 photons except
for the highest-energy bin which contains 15 photons. We used CIAO~2.1.3
Science Threads for the background subtraction and response matrix generation.
The spectrum was fitted between 0.3 and 6.7 keV (only 2 photons were detected
above this range) with a model combining a power law and a
neutron star hydrogen atmosphere model as calculated for the non-magnetic
case by Zavlin, Pavlov \& Shibanov (1996). Standard absorption by cold
interstellar matter of cosmological abundances was applied following Morrison
\& McCammon (1983). Due to the limited statistical quality, the fit errors
for the atmosphere temperature and radius are reasonably constrained only if
the mass of the neutron star, the distance, the interstellar hydrogen column
density $N_{\rm H}$ and the power law photon index are fixed. We did so at
values of 1.4~M$_\odot$ (for a 'canonical' neutron star), 8.5~kpc (this is well
determined), $N_{\rm H}=8.2\times10^{21}$~cm$^{-2}$ and 1.44 (these two
values were determined from the 1998 outburst; In 't Zand et al. 1999),
respectively. $N_{\rm H}$ has an uncertainty of roughly 40\%. The only three
parameters left free are the neutron star radius, the atmosphere temperature
and the power law normalization. The resulting fit is good, with $\chi^2_\nu=0.758$ (for $\nu=9$ degrees of freedom). The neutron star parameters are strongly
dependent. The fitted temperature is $92_{-33}^{+172}$~eV and the radius
$17_{-12}^{+31}$~km (90\% confidence). The radius to an observer at infinity is
$19_{-8}^{+29}$~km. A graph of the spectrum is presented in
figure~\ref{figspectrum} while in figure~\ref{figcontour} the 90\% confidence
region in temperature/radius space is presented. If $N_{\rm H}$ is left free,
it converges to $8.2^{+2.0}_{-1.3}\times10^{21}$~cm$^{-2}$ and the range of
allowed NS radii surpasses the 100 km mark. This upper radius margin is also
quite sensitive to the power law index which is poorly constrained by
the data.

The inferred radius is consistent with that expected for a neutron star
(e.g., Lattimer \& Prakash 2001). We note that the model implies an intrinsic
radius that is 20\% larger if the NS mass is 1.8 instead of 1.4~M$_\odot$.
Fits with some other models are nearly as good (table~\ref{goodness}), so the
data are not discriminate in that sense. The absorbed 0.5-7 keV flux is
$(5.9\pm0.4)\times10^{-14}$~erg~s$^{-1}$cm$^{-2}$, the unabsorbed
flux $(2.0\pm0.6)\times10^{-13}$~erg~s$^{-1}$cm$^{-2}$ (the large error
reflects the uncertainty in $N_{\rm H}$). The power law contribution
in the latter flux is 20$\pm$6\% (90\% confidence); this is only 8\% 
in 0.5 to 2 keV. The bolometric unabsorbed luminosity of the
NS atmosphere is $(2.1\pm0.8)\times10^{33}$~erg~s$^{-1}$ (the contribution
from the uncertainty in the distance is small).

%==============================================================================
\section{Discussion}
\label{sectiondis}

Pooley et al. (2001) find that there are 4 or 5 likely quiescent LMXB
transients in NGC~6440. Similar conclusions are currently being 
reached from {\em Chandra\/} observations of other globular clusters such as
47~Tuc (Grindlay et al. 2001a), Liller~1 (Homer et al. 2001), NGC~6652
(Heinke, Edmonds \& Grindlay 2001), $\omega$~Cen (Rutledge et al. 2001)
and NGC~6397 (Grindlay et al. 2001b).
Furthermore, for the first time two bright low-mass X-ray binaries have been
identified in a single cluster (M15; White \& Angelini 2001). 
Our observations show unequivocally that the second and third transient
outburst in NGC 6440 (in 1998 and 2001) are from the same object. 
Due to a limited positional accuracy of the source of the 1971 outburst,
it is not clear whether the same source was responsible for the 1971 and
the 1998/2001 outbursts. At first, indirect evidence suggested that the
source is different, because the duration of 1.5 to 3 months and the peak flux
of 250~mCrab deviate substantially from the values for the 1998 outburst
and for the initial phase of the 2001 outburst. However, at the time of this
writing the 2001 outburst shows activity for already 3 months and a peak flux
in excess of about 0.2 Crab units, according to ASM and WFC measurements. 

The recurrence time between the two most recent outbursts
is unexpectedly short. This may mean that the source has become
more active, or alternatively that earlier outbursts have been
missed. It is not uncommon for a previously quiet source, to
become active in a series of transient outbursts (e.g. GRO~J1744-28
and GRO~J1655-40). However, the 1998 outburst of the transient 
in NGC6440 was so faint and short, that similar outbursts may
easily have passed undetected. We note that the RXTE ASM and
BeppoSAX WFCs became operational only in 1996. Similar coverage as the ASM
in the future should resolve this issue.

This is the first time that an optical counterpart has firmly been established
for an X-ray transient in a globular cluster. Furthermore, it is the first
time that a quiescent neutron star has been identified without source
confusion in a globular cluster. These identifications are 
important because they present the case of a quiescent neutron star
transient in the optical and X-rays for which an accurate distance is known.

The spectral study of quiescent neutron stars is foremost interesting, as
Brown et al. (1998) pointed out, because it may result in an
accurate enough neutron star radius determination to be able to constrain the
equation of state of matter in the star's core. Measuring the radius to an
accuracy of about 0.5~km can exclude 50\% of the proposed equations of states
(Lattimer \& Prakash 2001). For neutron stars in globular clusters as compared
to those in the Galactic disk, one source of uncertainty is almost eliminated
--- the distance. The distance to NGC~6440 is accurate to 5\%. If other
uncertainties (in the spectral parameters) had been at least an order
of magnitude lower, the
prospects for constraining the equation of state for the transient in
NGC~6440 would have been good. However, this is not the case. This qNS would
have to be observed for a few million seconds with {\rm Chandra} before
it becomes worthwhile to consider this analysis, and that is without
considering uncertainties in the spectral models.

Given a peak bolometric luminosity between $4\times10^{37}$ and
$1.6\times10^{38}$~erg~s$^{-1}$ (or 50 to 200~mCrab in 2-10 keV given the
broad-band spectrum measured by In~'t~Zand et al. 1999),
an e-folding decay time between 5~d (Verbunt et al. 2000) and 30~d, and
assuming that during accretion 100\% of the gravitational
potential energy is released through radiation, the recurrence time of
2.95~yr implies a time-averaged accretion rate of
$<\dot{M}>\approx2\times10^{-11}$ to $4\times10^{-10}~M_\odot$~yr$^{-1}$.
Following Brown
et al.\ (1998), the quiescent time-averaged luminosity due to thermal 
emission from the cooling NS is predicted to be between
$1\times10^{33}$ and $2\times10^{34}$~erg~s$^{-1}$. Our measurement of
$2.1\times10^{33}$~erg~s$^{-1}$ is consistent with this rough prediction,
also if less than 100\% of the liberated potential energy
is released as radiation.

%------------------------------------------------------------------------------
\acknowledgements
We thank Harvey Tananbaum for the willingness to schedule the {\em Chandra\/}
observation at short notice, Yousaf Butt for help in planning the {\em
Chandra\/} observation, Olivier Hainaut
for carrying out the optical observations, Rob van der Meer for help in the
ACIS data analysis, Slava Zavlin and George Pavlov for providing their
neutron star atmosphere models in electronic form, and Jaap Schuurmans for
software support. JZ acknowledges financial support from the Netherlands
Organization for Scientific Research (NWO), MHvK through a fellowship of the
Royal Netherlands Academy of Sciences (KNAW), DP through a National Science
Foundation Graduate Fellowship, WL from NASA, and RW from NASA through a
Chandra fellowship.

%==============================================================================

%==============================================================================

\newpage

\begin{table}[t]
\caption[]{Goodness of fit for a variety of spectral models
to the 12-channel data between 0.3 and 6.7 keV.\label{goodness}}
\begin{tabular}{lllr}
\hline\hline
Model$^\dag$    & free                            & $\chi^2_\nu$ & $\nu$ \\
                & params.                         &              &       \\
\hline
wa*nsa          & $T,R$                           &   2.74  &10 \\
wa*(nsa+po)     & $T,R$,po-norm                   &   0.758 & 9 \\
wa*bb           & $T$,bb-norm                     &   3.00  &10 \\
wa*(bb+po)      & $T$,bb-norm,po-norm             &   0.987 & 9 \\
wa*diskbb       & $N_{\rm H}$,$T_{\rm in}$,norm   &   1.94  & 9 \\
wa*diskbb       & $T_{\rm in}$,norm               &   2.44  &10 \\
\hline\hline
\end{tabular}

\noindent
$^\dag$wa = interstellar absorption following Morrison
\& McCammon (1983), nsa = NS H-atmosphere emission according to Zavlin et
al. 1996, po = power law, bb = black body, diskbb = disk black body
according (Mitsuda et al. 1984)
\end{table}

\end{document}